\documentclass[twocolumn]{aastex631}

\usepackage{gensymb}

\newcommand\RA{$19^{h}41^{m}17.5^{s}$}
\newcommand\DEC{$+40^{\degree}11^{'}47^{"}$}

\newcommand\ourage{2.428^{+0.0040}_{-0.0045}}
\newcommand\ourdist{2.419^{+0.0008}_{-0.0005}}
\newcommand\ourmetallicity{-0.035^{+0.0026}_{-0.0029}}
\newcommand\ourav{556.5^{+2.0}_{-2.1}}

\newcommand\numbinaries{338} 
\newcommand\numbinariesqlim{250}
\newcommand\numclustermembers{1342} 
\newcommand\binaryfrac{0.186} 
\newcommand\binaryfracerr{0.012}
\newcommand\upperlim{14.85}
\newcommand\lowerlim{19.5}
\newcommand\clustermin{0.1}
\newcommand\qmin{0.5}
\newcommand\binaryfracpval{1.50$\times10^{-3}$}
\newcommand\chisquaredpval{0.204}

\newcommand\binaryvssinglepval{1.47$\times10^{-3}$}
\newcommand\binarykspval{2.49$\times10^{-4}$}
\newcommand\singlekspval{1.99$\times10^{-4}$}

\newcommand\totalmembers{2632}
\newcommand\totalbinaries{777}

\begin{document}

\title{Investigating Mass Segregation of the Binary Stars in the Open Cluster NGC 6819}

\author[0000-0003-3695-2655]{Claire Zwicker}
\affiliation{Illinois Institute of Technology, 10 West 35th Street Chicago, IL 60616, USA}

\author[0000-0002-3881-9332]{Aaron M. Geller}
\affiliation{Center for Interdisciplinary Exploration and Research in Astrophysics (CIERA) and Department of Physics and Astronomy, Northwestern University, 1800 Sherman Avenue, Evanston, IL 60201, USA}

\author[0000-0002-9343-8612]{Anna C. Childs}
\affiliation{Center for Interdisciplinary Exploration and Research in Astrophysics (CIERA) and Department of Physics and Astronomy, Northwestern University, 1800 Sherman Avenue, Evanston, IL 60201, USA}

\author[0009-0001-9841-0846]{Erin Motherway}
\affiliation{Embry-Riddle Aeronautical University, Department of Physical Sciences, 1 Aerospace Blvd, Daytona Beach, FL 32114, USA}

\author[0000-0002-5775-2866]{Ted von Hippel}
\affiliation{Embry-Riddle Aeronautical University, Department of Physical Sciences, 1 Aerospace Blvd, Daytona Beach, FL 32114, USA}

\begin{abstract}
We search for mass segregation in the intermediate-aged open cluster NGC 6819 within a carefully identified sample of probable cluster members.  Using photometry from the Gaia, 2MASS, and Pan-STARRS surveys as inputs for a Bayesian statistics software suite, BASE-9, we identify a rich population of (photometric) binaries and derive posterior distributions for the cluster age, distance, metallicity and reddening as well as star-by-star photometric membership probabilities, masses and mass ratios (for binaries).  Within our entire sample, we identify \totalmembers\  cluster members and \totalbinaries\  binaries.  We then select a main-sequence ``primary sample'' with \upperlim\ $<G<$ \lowerlim\ containing \numclustermembers\ cluster members and \numbinariesqlim\ binaries with mass ratios $q>$\ \qmin, to investigate for mass segregation. Within this primary sample, we find the binary radial distribution is significantly shifted toward the cluster center as compared to the single stars, resulting in a binary fraction that increases significantly toward the cluster core. Furthermore, we find that within the binary sample, more massive binaries have more centrally concentrated radial distributions than less massive binaries. The same is true for the single stars. We verify the expectation of mass segregation for this stellar sample in NGC 6819 through both relaxation time arguments and by investigating a sophisticated $N$-body model of the cluster.
Importantly, this is the first study to investigate mass segregation of the binaries in the open cluster NGC 6819.
\end{abstract}

\keywords{Binary stars (154) --- Open star clusters (1160) --- Relaxation time (1394) --- Bayesian statistics (1900) --- N-body simulations (1083)}

\section{Introduction} \label{sec:intro}

NGC 6819 is an intermediate-aged open cluster and home to a rich binary population \citep{Hole_et_al._2009, 2014AJ....148...38M, 2022MNRAS.512.3992C}.  With an age of about 2.5 Gyr \citep[e.g.,][]{Kalirai_et_al., 2011ApJ...729L..10B, 2013AJ....146...58J, 2013ApJ...762...58S,Yang_et_al.2013, Ak_et_al._2016, 2016AJ....151...66B}, NGC 6819 has persisted through $\sim$5-10 half-mass relaxation times \citep[e.g.,][]{Kalirai_et_al., Kang_&_Ann, Karatas_et_al.2023} and therefore should be relatively dynamically relaxed.  One product of two-body relaxation is an expectation that the more massive single and binary stars in the cluster will occupy a more centrally concentrated spatial distribution than those with lower mass \citep[e.g.,][]{King1962, 1997MNRAS.286..709G}.

\begin{figure*}[!ht]
\plotone{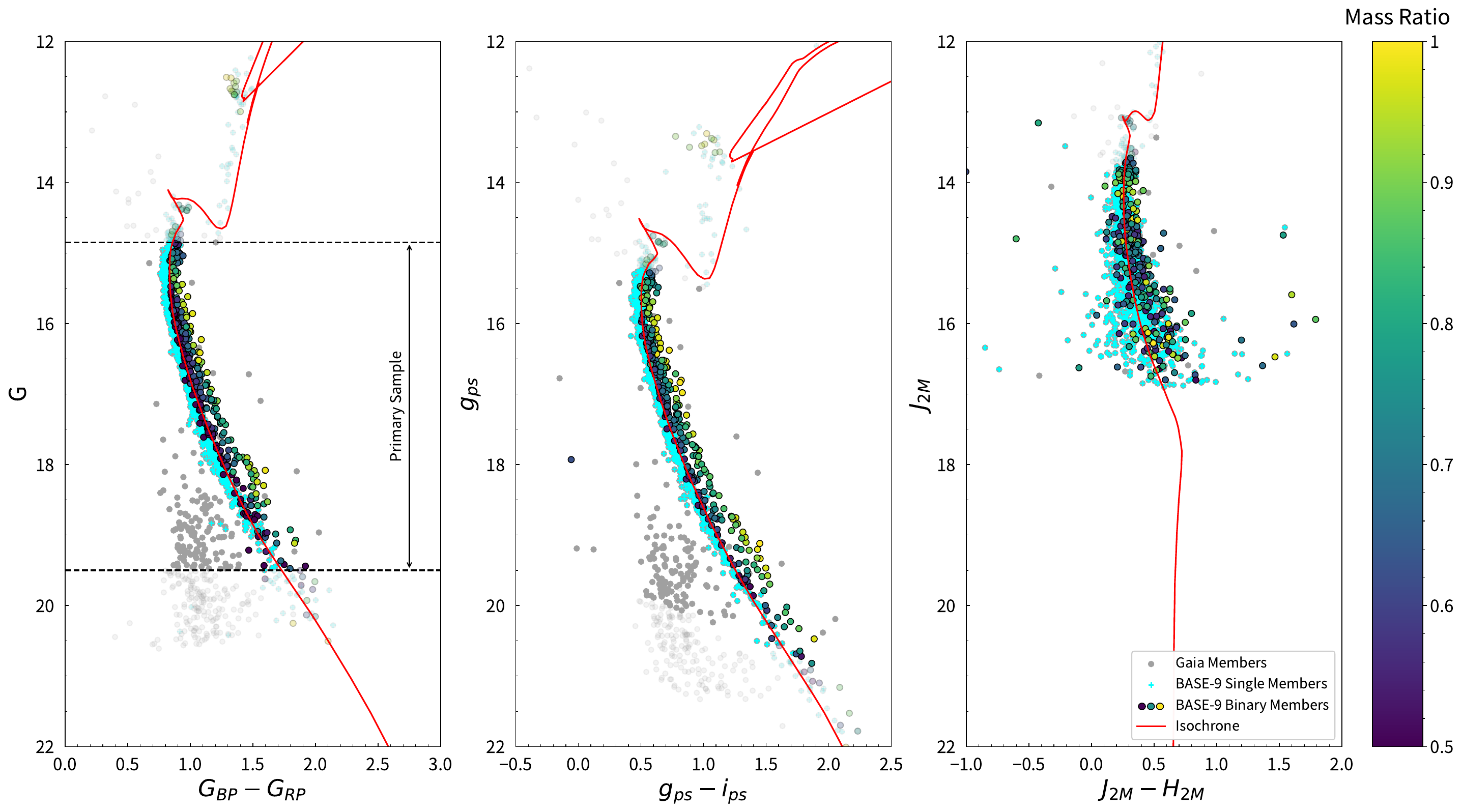}
\caption{Color-magnitude diagrams (CMDs) of NGC 6819 with photometry from Gaia (left), Pan-STARRS (center), and 2MASS (right). We plot BASE-9 members in colored symbols, with singles in cyan plus symbols and identified binaries in points colored by their median mass ratio, as indicated by the colorbar on the right. Additional Gaia members that did not pass the BASE-9 member cut are plotted as gray circles. We show our primary sample selection, made using the Gaia $G$ filter, with horizontal dashed lines in the Gaia (left) CMD. In all panels, stars within the primary sample are plotted with higher opacity than stars outside of this primary sample. Also in all panels we include a PARSEC isochrone generated using the median values from the posterior distributions resulting from our BASE-9 analysis. 
\label{fig:CMD}}
\end{figure*}

Several previous studies have found evidence that NGC 6819 is mass segregated. \cite{Kalirai_et_al.} compared luminosity functions in radial annuli and found the central annulus to be weighted toward brighter (higher-mass) stars as compared to an almost flat luminosity function observed for the entire cluster. They interpret this finding as evidence for dynamical evolution causing higher mass stars to sink to the inner regions of the cluster. \cite{Kalirai_et_al.} also examined mass functions across eight different annuli, finding that the slope changes from positive (more higher-mass stars) to negative (more lower-mass stars) with increasing radial distance from the cluster center, consistent with the expectations of dynamical evolution.  
\cite{Kang_&_Ann} constructed cumulative distribution functions (CDFs) of the number of stars with respect to the distance from the cluster center in bins of magnitude, finding that brighter stars have more centrally concentrated radial distributions. They also calculated ``half number radii'' for their brightest and faintest magnitude bins confirming this result.
\citet{Yang_et_al.2013} also constructed mass functions in radial annuli and confirm the cluster to be mass segregated.

Since these studies were published, we now have access to precise kinematic and parallax information for stars in NGC 6819 from Gaia \citep{2016A&A...595A...1G, 2023A&A...674A...1G}.  These data are extremely helpful in separating field stars from cluster members \citep{2018A&A...618A..93C}.  Indeed \citet{Karatas_et_al.2023} used Gaia EDR3 data to isolate cluster members in NGC 6819 (and other clusters) to study the cluster structure, dynamics, mass segregation and the Galactic orbit.  Reliably distinguishing cluster members is particularly important for mass segregation studies; as the cluster itself becomes sparser farther from the cluster center, field-star contamination becomes more of a concern and may impact the derived luminosity and mass distributions and radial profiles.  In this study, we use Gaia radial-velocity, proper-motion and parallax measurements (where available) along with a photometric membership analysis to identify probable cluster members and limit the effects of field-star contamination.

Importantly, here we focus on the binary stars.  For a given primary-star mass, a binary (containing two stars) is more massive than the corresponding single star, and therefore mass segregation effects are expected to move the binaries into a more centrally concentrated radial distribution than the singles.  On the other hand, close gravitational encounters between binaries and other cluster members, which happen preferentially in the dense cluster core, can also destroy binaries \citep[e.g.][]{1975AJ.....80..809H}.  Models predict that by a dynamical age similar to that of NGC 6819, mass segregation will dominate over binary destruction \citep{2013ApJ...779...30G}. Indeed studies of older rich open and globular clusters have found the binaries to be more centrally concentrated than the single stars \citep[e.g.,][]{Geller_et_al._2008, 2012A&A...540A..16M, 2015AJ....150...97G, 2021AJ....162..264J}.  

However, observational evidence of mass segregation in clusters around the age of NGC 6819 is mixed.  \citet{2018MNRAS.473..849D} characterized mass segregation in 1276 OCs and found that only 14\% of these clusters showed significant evidence of mass segregation.  Of these 1276 OCs, nine have cluster ages that are within $\pm 50$ Myr of the age of NGC 6819, and only four of these show some evidence of mass segregation.  The ubiquity of mass segregation for binary stars in particular becomes even less certain, in part because identifying unresolved binaries and mapping their radial distribution is a challenge.  For example, \citet{2021AJ....162..264J} studied multiple OCs and found that $>$50\% of the OCs in their study do not show significant evidence of mass segregation of the binary stars relative to the single stars. More specifically, of the 23 OCs they studied, three have ages within $\sim$1 Gyr of NGC 6819 (though all are younger than NGC 6819), and of these, two show some evidence of mass segregation of the binaries; however one of these has a mass much smaller than NGC 6819.  For the first time, here we investigate NGC 6819 for evidence of mass segregation within the binary population and compare our observational results to an $N$-body model of the cluster.  

In Section \ref{sec:methods} we define our primary sample, and how we determine cluster membership and global cluster parameters from our  data. In Section \ref{sec:results} we study the radial distributions of the binary and single stars in the cluster. In Section \ref{sec:Nbodyresults} we present a comparison to a direct $N$-body model of a cluster like NGC 6819. Finally, in Section \ref{sec:discussion} we provide a discussion and conclusions.

\section{Cluster Membership, Global Parameters and our Primary Stellar Sample}\label{sec:methods}

We follow a nearly identical procedure to \citet{2024ApJ...962...41C} to prepare a sample for analysis of mass segregation in NGC 6819.  In short, first we download Gaia \citep{2023A&A...674A...1G}, Pan-STARRS \citep{2020ApJS..251....6M} and 2MASS \citep{2006AJ....131.1163S} data for all stars within the cluster's effective radius of $0.42\degree$ (\citealt{2024ApJ...962...41C}, equivalent to $\sim$4.5 core radii as derived from a \citealt{1962AJ.....67..471K} model fit to our data) from the center of NGC 6819,  $\alpha$ = \RA \ and $\delta$ = \DEC \citep{2013AJ....146...43P}.  We use the Gaia DR3 radial-velocity, proper-motion and parallax distributions to derive priors on the cluster membership for each star (a given star's membership prior is calculated based on the distance that the star's Gaia kinematics and parallax values are measured away from the cluster's mean measurements; see \citealt{2024ApJ...962...41C} for more details).  We follow the same procedure as \citet{2024ApJ...962...41C} to use the Bayesian Analysis for Stellar Evolution with Nine Parameters (BASE-9) statistics software suite \citep{2006ApJ...645.1436V, 2009AnApS...3..117V, 2016ascl.soft08007R} and the PARSEC isochrone models \citep{2012MNRAS.427..127B} to derive the posterior distributions for the global cluster parameters (cluster age, distance, reddening and metallicity) for NGC 6819. \citealt{2024ApJ...962...41C} find the following median values and 1$\sigma$ uncertainties for the posterior distributions of each global cluster parameter: an age of $\ourage$ Gyr, a distance of $\ourdist$ kpc, [Fe/H] = $\ourmetallicity$ dex, and  $E(B-V) =$ $\ourav$ mmag. We show isochrones using these median values in Figure~\ref{fig:CMD}.

\begin{figure}[!t]
\plotone{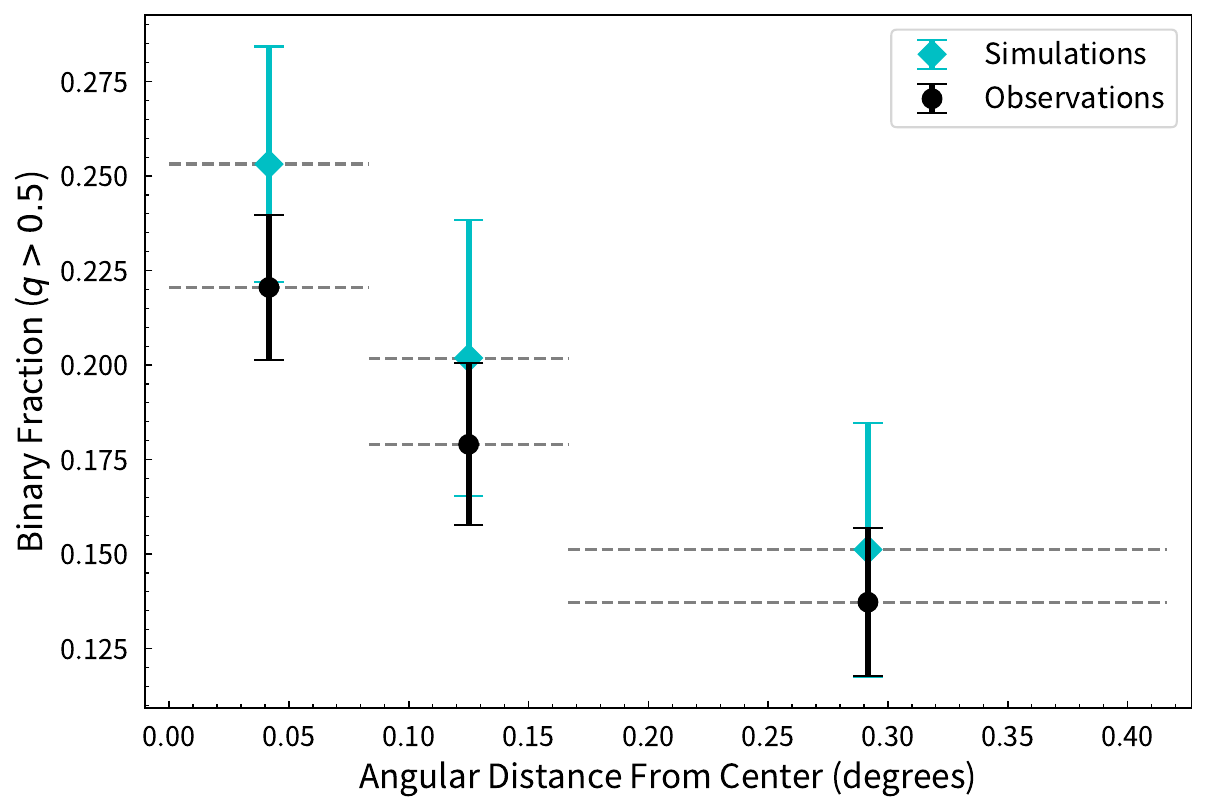}
\caption{Binary fraction for stars in our observed primary sample compared to similar stars in our NGC 6819 $N$-body model (see Section~\ref{sec:Nbodyresults}), with respect to angular distance from the cluster center. The observational data is plotted as black circles while the simulation data is plotted in cyan diamonds. For each observational data point, we divide the number of binaries with $q > \qmin$ by the number of total objects (including all single and binary members, regardless of $q$) in our primary sample within the bin defined by the horizontal dashed line.  We show the 1$\sigma$ uncertainties (of both data sets) in each bin with vertical error bars. The first two bins have equal width, of about one core radius.  The width of the third bin was chosen to contain approximately the same number of stars as the previous bin (as the number of stars decreases dramatically toward the edge of the cluster). We follow the same bins and sample selection for the NGC 6819 model; however here the uncertainties show the $1\sigma$ width of the distribution of binary fractions in all simulations in our model. The binary fraction decreases with increasing radius from the cluster center for both the observed and simulated data.
\label{fig:binaryfrac}}
\end{figure}

Next, we use BASE-9 to derive star-by-star posterior distributions for each star's (primary) mass, mass ratio ($q$, if a binary), and photometric membership probability.  For input here we use a more selective sample than \citet{2024ApJ...962...41C}; we choose to only include stars whose Gaia membership prior is $>\ $\clustermin.  We impose these limits, after some exploration of the data, in an attempt to further remove field star contaminants in the very rich field of NGC 6819. We will refer to the stars that satisfy these criteria as ``Gaia members'', and we use these Gaia members as inputs for this second step in our BASE-9 analysis.  BASE-9 then provides a photometric membership estimate for each star; we follow \citet{2024ApJ...962...41C} and require a minimum median photometric membership value $\geq 0.01$ (in addition to the Gaia membership limit) to consider a star a cluster member. Again, this limit was found through experimentation, with the goal to limit field-star contamination while also not excluding a significant amount of true members.  Stars in NGC 6819 down to a Gaia $G$ magnitude of 20.8 that pass both our Gaia and BASE-9 membership limits are shown in Figure~\ref{fig:CMD}.  We  use this sample of members in our subsequent analysis.  Finally, we follow \citet{Cohen_et_al.} and identify BASE-9 members as binaries if their posterior distribution of secondary mass has a median value that is $\geq$ $3\sigma$ above zero.

This procedure results in a sample of \totalmembers\ BASE-9 members.  We show these stars, along with PARSEC isochrones, for different photometric filter combinations in Figure~\ref{fig:CMD}. All underlying data for this paper has been published to Zenodo \citep{Childs_Zenodo}.

\begin{figure}[!t]
\plotone{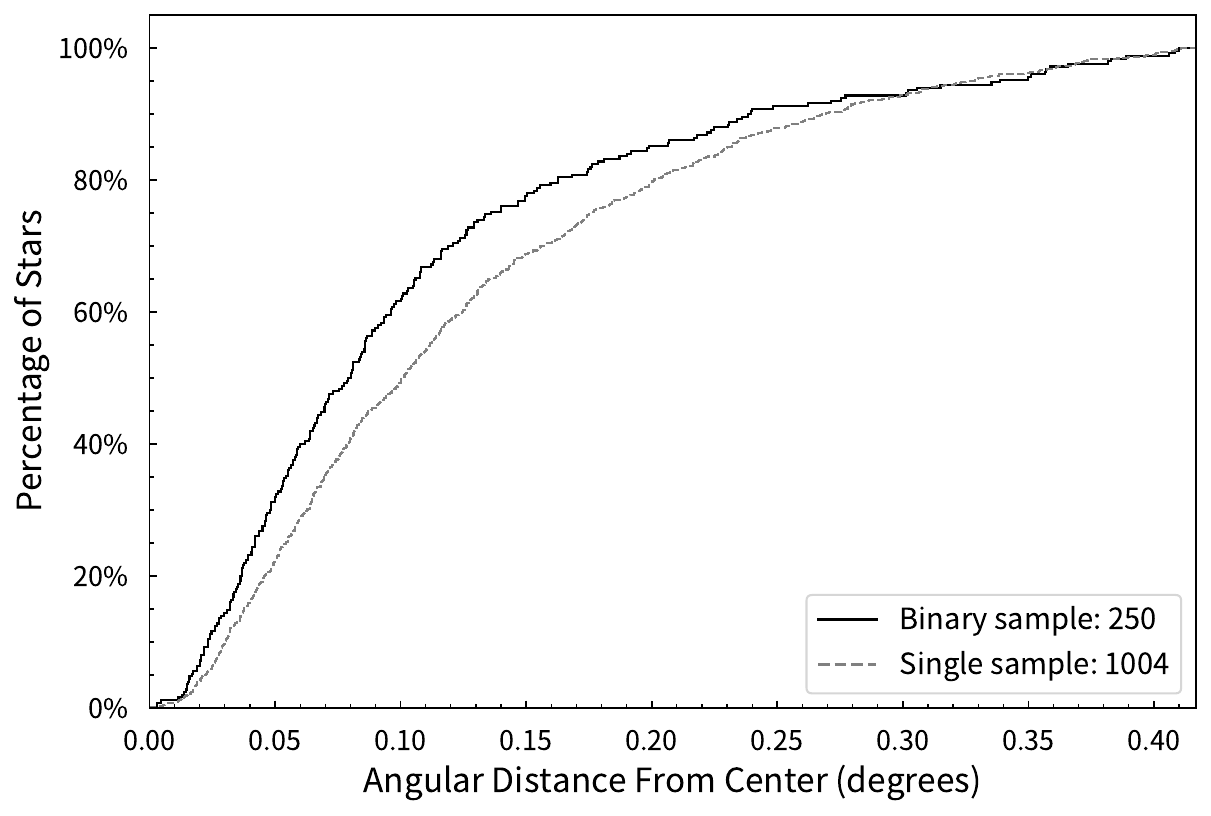}
\caption{CDF of all NGC 6819  $q>\ $\qmin\ binary (solid black line) and single (dashed gray line) BASE-9 members in our primary sample with respect to angular distance from the cluster center. The radial distribution of the binaries is shifted toward the cluster center as compared to that of the single stars.
\label{fig:CDFbinaryvssingle}}
\end{figure}

\subsection{Primary Sample Selection}\label{sec:primary_sample}
Though we use all available BASE-9 members within $0.42\degree$ of the cluster center to determine global cluster parameters, we limit our sample for our subsequent analysis to only contain main-sequence stars.  Specifically, we define a ``primary sample'' that spans from $\upperlim\ <G< \lowerlim$.  We chose the bright limit to remove stars near and above the turnoff and the faint limit to remove stars that have large enough uncertainties in the optical bands to hinder our ability to detect photometric binaries.  We target main-sequence stars for this analysis because the giants may have lost a substantial amount of mass, which could modify their radial distribution.  In total we find \numclustermembers\ stellar systems in our primary sample with \numbinaries\ binaries (over all $q$).  For $q > 0.5$ our binary sample is nearly complete \citep[see]{Cohen_et_al., 2024ApJ...962...41C}.  However our ability to detect binaries becomes incomplete at low $q$, as these binaries have minimal separation from the isochrone, making them difficult for BASE-9 to detect. Therefore we restrict our analysis of the binaries in our primary sample to the \numbinariesqlim\ binaries with  $q >$ \qmin.  This translates to a binary fraction for $q>$\ \qmin\ of $\binaryfrac\ \pm\ \binaryfracerr$.

\begin{figure*}[!ht]
\plotone{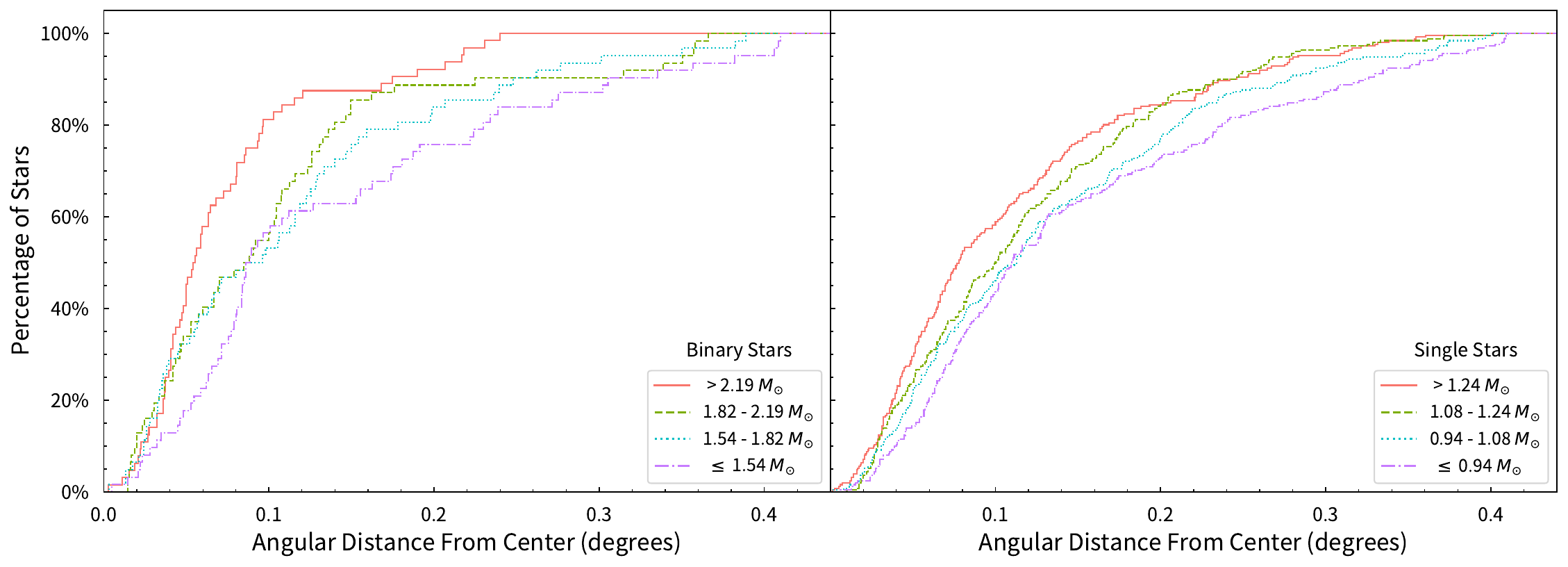}
\caption{CDFs of the $q>$\ \qmin\ binary-star (left) and single-star (right) populations separated into mass bins.  For each respective sample, we attempt to choose bins with equal sample sizes; for the binaries the three lower-mass bins contain 62 systems while the highest-mass bin contains 64 systems, and for the singles all bins contain 251 stars. Both populations are plotted with respect to angular distance from the cluster center, extending to $0.42\degree$.  Both the binary and single stars show strong evidence of mass segregation. 
\label{fig:massCDFs}}
\end{figure*}

\section{Radial Distributions of the Single and Binary Star Populations}\label{sec:results}

In Figure \ref{fig:binaryfrac} we plot the binary fraction of our observed primary sample compared to a sophisticated $N$-body model of NGC 6819 $N$-body (see Section~\ref{sec:Nbodyresults}) with respect to distance from the cluster center. Here we discuss the observations and save a discussion of the $N$-body model until Section~\ref{sec:Nbodyresults}.  We see an overall decrease in the binary fraction as distance from the cluster center increases. Moreover, a two-sided Z-test between the first and third bins of the observations returns a significant distinction, with a $p$-value of \binaryfracpval.  

In Figure \ref{fig:CDFbinaryvssingle} we present CDFs of all NGC 6819 single (dashed gray line) and $q>\ $\qmin\ binary (solid black line) BASE-9 members in our primary sample as a function of distance from the cluster center. A two-sample Kolmogorov-Smirnov (K-S) test between both populations returns a p-value of \binaryvssinglepval.  We therefore conclude that the binaries and single stars are drawn from distinct parent populations. The binaries are centrally concentrated with respect to the single stars in our primary sample. 

In Figure \ref{fig:massCDFs} we investigate mass segregation signatures within the single and binary star populations respectively, divided into four bins of increasing mass. 
On the left we show the binary population, and on the right we show the single-star population.  Visually it is clear that higher-mass single or binary stars in the cluster are more centrally concentrated.  A two-sample K-S test comparing the lowest and highest mass bins for the single stars returns a p-value of \singlekspval, and a similar test for the binary star samples returns a p-value of \binarykspval. Thus we find that the more massive stellar systems, for both the single and binary samples respectively, are more centrally concentrated than the least massive systems.  

\subsection{Relaxation timescales for NGC 6819\label{subsec:dynamic}}
 
The half-mass relaxation time, $t_{rh}$, is a characteristic timescale for mass segregation and cluster relaxation. We calculate the half-mass relaxation time for NGC 6819 following \citet{Binney_Tremaine} and \citet{1969ApJ...158L.139S}:

\begin{equation}
\label{eq1}
    t_{rh}= \frac{0.17N}{\ln(\lambda N)}\sqrt{\frac{r_{h}^3}{GM}}
\end{equation}

where $N$ is the number of cluster members, $\lambda$ is a constant we assume to be 0.1  \citep{1994MNRAS.268..257G}, $r_{h}$ is the half-mass radius, $M$ is the total mass of the cluster, and $G$ is the gravitational constant.  To estimate the half-mass relaxation time, we use all cluster members in our sample, no longer limited by the primary sample selected above.  We estimate a half-mass radius both by investigating the radial mass distribution (which we assume is incomplete) and using a simple conversion from core radius \citep{Heggie_&_Hut2003}, and find $r_h$ to be between roughly 5 and 10 pc, assuming a distance of $\ourdist$ kpc and accounting for projection effects (by multiplying our observed projected half-mass radius by a factor of 4/3, following \citealt{1987degc.book.....S}).  For our calculation we use  $r_h = 7.5 \pm 2.5\ $pc (de-projected). Within our sample we find $M \sim 3100\ M_\odot$ and $N \sim 2800$ out to the cluster's effective radius. 
Previous literature reports NGC 6819 having a mass and number of members in the range $M \sim$ $2100-2600$ $M_\odot$ and $N \sim 1900-2900$ \citep{Kalirai_et_al.,Kang_&_Ann, Yang_et_al.2013}. Due to the varying methodologies employed to calculate the total cluster mass, each incomplete in its own way, we opt for a cautious approach, settling on an intermediate value of M = 2700 $\pm$ 600 $M_\odot$ and N = 2400 $\pm$ 400.  This results in $t_{rh}$ = 440 $\pm$ 230\ Myr.  This is somewhat larger than, though still consistent with previous values from the literature \citep{Kalirai_et_al., Kang_&_Ann, Yang_et_al.2013}. 
Given the age from our BASE-9 analysis of $\ourage$ Gyr, we find that NGC 6819 has survived more than five half-mass relaxation times. 

We can also estimate a mass-segregation timescale for stars of a given mass, following \citet{Spitzer_et_al.} :

\begin{equation}
\label{eq2}
    t_{seg}= \frac{\langle m \rangle}{m}t_{rh}
\end{equation}

where $\langle m \rangle$ is the average mass of an object in the cluster, $m$ is the mass of the object of interest, and $t_{rh}$ is the global half-mass relaxation time for the cluster, which we calculated above. 
Within our data, we find $\langle m \rangle \sim 1.26 M_\odot$ (which is likely an overestimate of the true $\langle m \rangle$, given that our cluster data is likely incomplete at the faint end).  For the analysis shown in Figure~\ref{fig:massCDFs}, even the lowest mass bin has a mass-segregation time of $\sim\ $4 times smaller than the cluster age (and the higher-mass bins have even shorter $t_{seg}$).  Thus, both the single and binary stars in the cluster are expected to have had ample time to mass segregate dynamically, as is readily apparent in the empirical results shown in Figure~\ref{fig:massCDFs}.

\section{Comparison to a Direct $N$-body Model of the Cluster}\label{sec:Nbodyresults}

In order to further investigate the theoretical expectations for mass segregation of the binaries and single stars in NGC 6819, we constructed an $N$-body model to approximate the cluster. 
In this section we describe the setup of the model, perform mock observations of the model, and compare those with the observations of the real cluster (presented above). 
 
 \subsection{$N$-body Model Setup}
 
Our goal is to create an $N$-body model that is reasonably similar to the cluster within the parameters that are important to this study (e.g., the cluster density, mass, age, and binary fraction).  We therefore began by investigating a large existing grid of $N$-body star cluster models that our group has used for various projects in the past \citep[e.g.][]{2017AAS...22924706F,2024AAS...24345819M}.  This grid was created using the \texttt{nbody6++gpu} code \citep{2003gnbs.book.....A,2015MNRAS.450.4070W} with updates to the initial binary parameters and the output files as described in \citet{2013AJ....145....8G, 2013ApJ...779...30G}.  In short these updates aim to produce an initial binary population that is roughly consistent with binaries observed in the Galactic field, across all spectral types \citep[in orbital parameters and mass ratio distributions and relative binary fraction;][]{2010ApJS..190....1R, 2013ARA&A..51..269D, 2017ApJS..230...15M}.  

The code then evolves the cluster, accounting for gravitational dynamics, stellar and binary (and higher-order) evolution, and other relevant astrophysics, and provides snapshots at regular intervals that contain stellar evolutionary and binary orbital parameters for each star (along with other summary information about the cluster).  Our existing grid of models samples a range in initial number of stars, half-mass radius, Galactocentric radius and metallicity.  All begin with the same initial binary population, consistent with the Galactic field.  We focus here on models that have a solar metallicity and are evolved within a standard Solar orbit in the Galactic potential.

We began by investigating models within this subset of the grid that reached the age of NGC 6819 by comparing the observed and simulated surface density profiles.  We found that a simulation with initially 20,000 stars and an initial half-mass radius of 2.6 pc produces approximately the observed number of stars and surface density profile at the age of NGC 6819.  Turning to investigate the binary fraction at the age of NGC 6819, we found that this simulation overestimated the number of binaries for the (approximately) solar-type stars that we have access to in the observations.  Interestingly, this suggests that NGC 6819 may have been born with a lower binary fraction than is observed in the Galactic field;  we return to this briefly in Section~\ref{sec:discussion}.  

By comparing the simulated and observed solar-type binary fractions at the age of NGC 6819, we estimated that we required a reduction in the initial binary fraction by 27\% in order to match the observations.  We then ran eight simulations with this updated binary population (without modifying the other initial parameters); each simulation was initialized with the same initial parameter distributions but with a different initial random seed to attempt to account for variations introduced by the random nature of dynamical systems (and also any numerical artifacts that may affect the result), as is standard procedure when modelling open clusters \citep[e.g.][]{2001ApJ...555..945K, 2009MNRAS.397.1577P, 2013AJ....145....8G}.   We will refer to this collection of $N$-body simulations as the ``NGC 6819 model'' in the subsequent analysis.          

\subsection{Comparison of the NGC 6819 Model and Observations}

\begin{figure}[!t]
\plotone{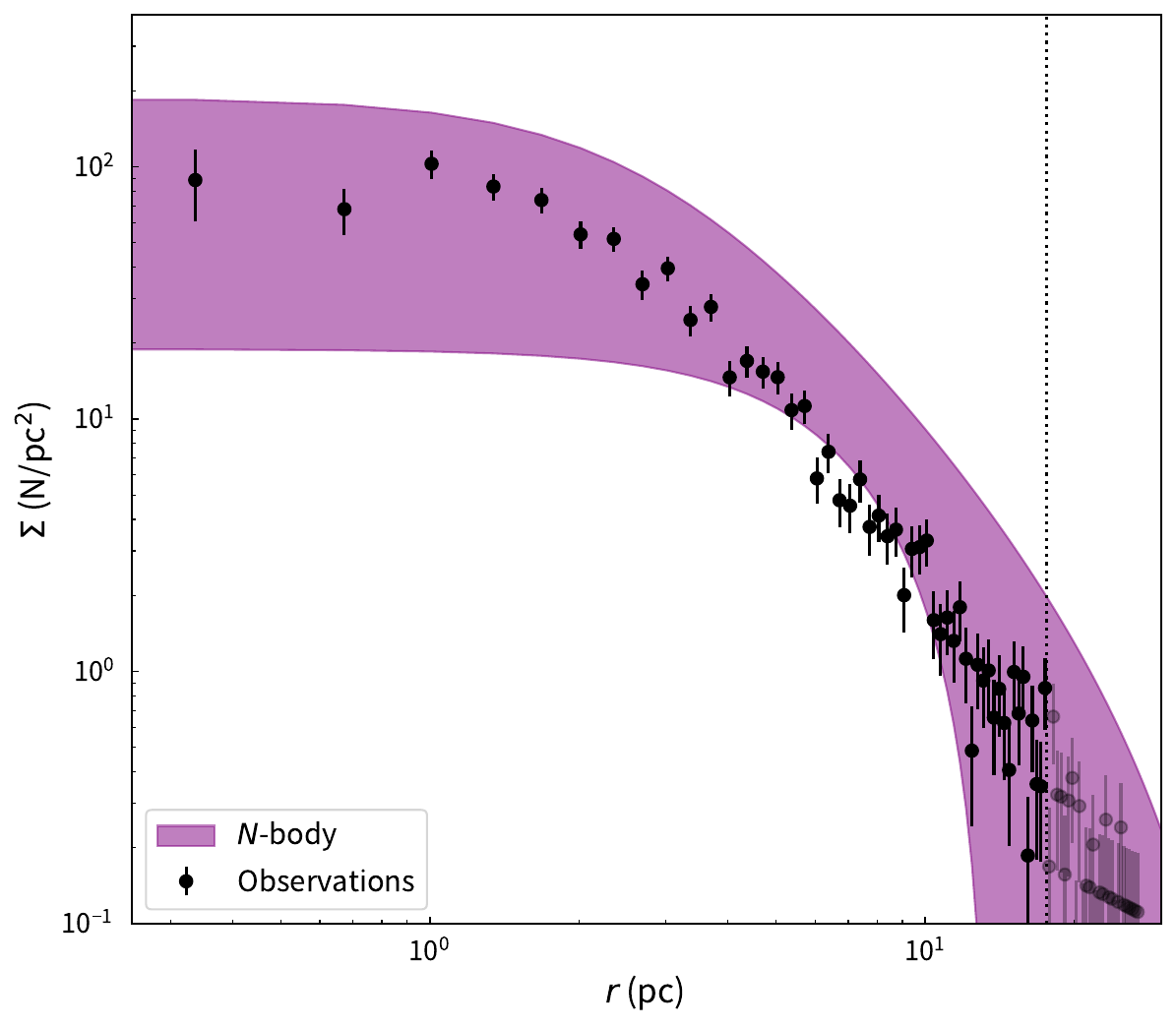}
\caption{Projected surface density radial profile of our observational sample of NGC 6819 compared to the NGC 6819 $N$-body model. The purple band shows the 3$\sigma$ bounds of \citet{King1962} models fit to the NGC 6819 data. The observational data of NGC 6819 is plotted in black points, with vertical error bars indicating the $1\sigma$ uncertainties. The gray, dotted vertical line indicates NGC 6819's effective radius \citep{2024ApJ...962...41C}.  Our analysis of both the model and observations only extends to the effective radius. 
\label{fig:surfdensity}}
\end{figure}

\begin{figure}[!t]
\plotone{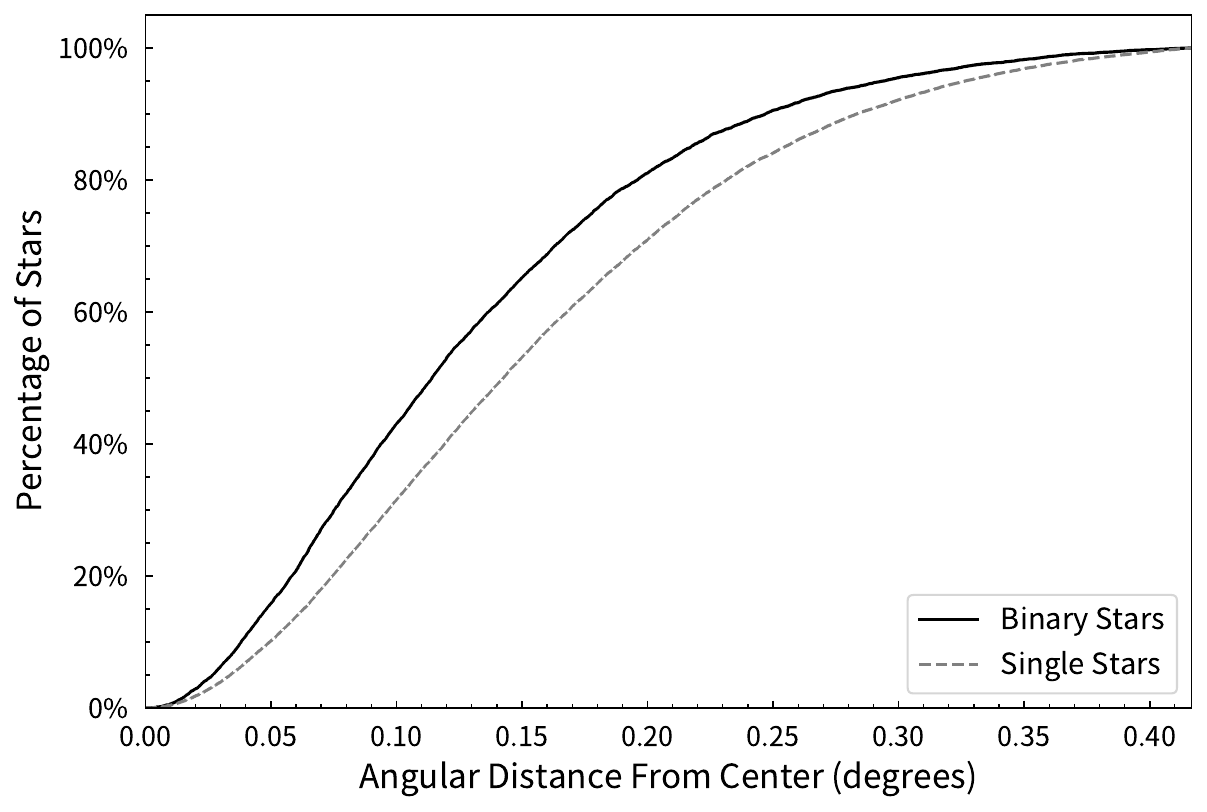}
\caption{CDF of binary (solid black line) and single (dashed gray line) stars within the NGC 6819 model as a function of angular distance from the cluster center. We include only stars that would fall within the primary sample, and show only binaries with $q > 0.5$ (as we also do for the observations in Figure \ref{fig:CDFbinaryvssingle}). The binary stars within the NGC 6819 model are more centrally concentrated than the single stars, which corresponds with the observational result shown in Figure \ref{fig:CDFbinaryvssingle}.
\label{fig:NbodyCDFbinaryvssingle}}
\end{figure}

\begin{figure*}[!ht]
\plotone{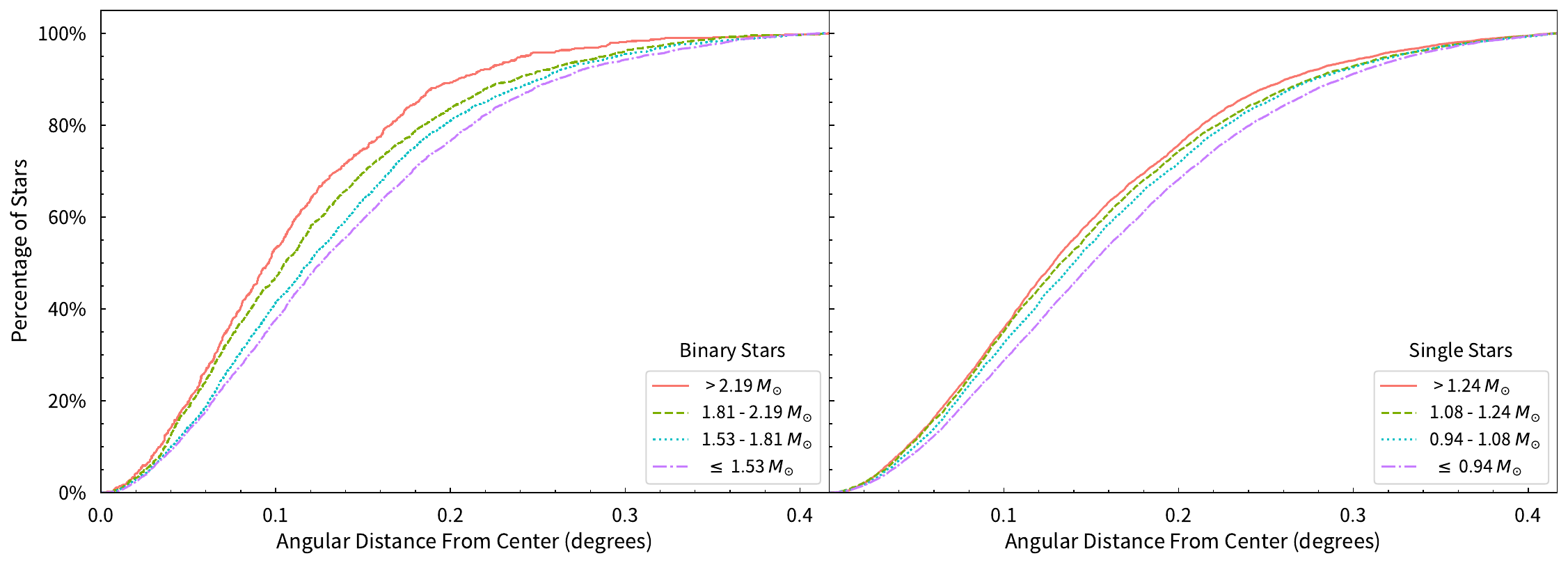}
\caption{CDFs of the binary-star (left) and single-star (right) populations within the NGC 6819 model separated into mass bins. The mass bins are constructed with the same ranges used in Figure \ref{fig:massCDFs}, for each respective subplot. Both populations are plotted with respect to angular distance from the cluster center. The binary-star and single-star populations within the NGC 6819 model show clear evidence of mass segregation, which corresponds with the observational result shown in Figure \ref{fig:massCDFs}.
\label{fig:NbodymassCDFs}}
\end{figure*}

\begin{figure}[!ht]
\plotone{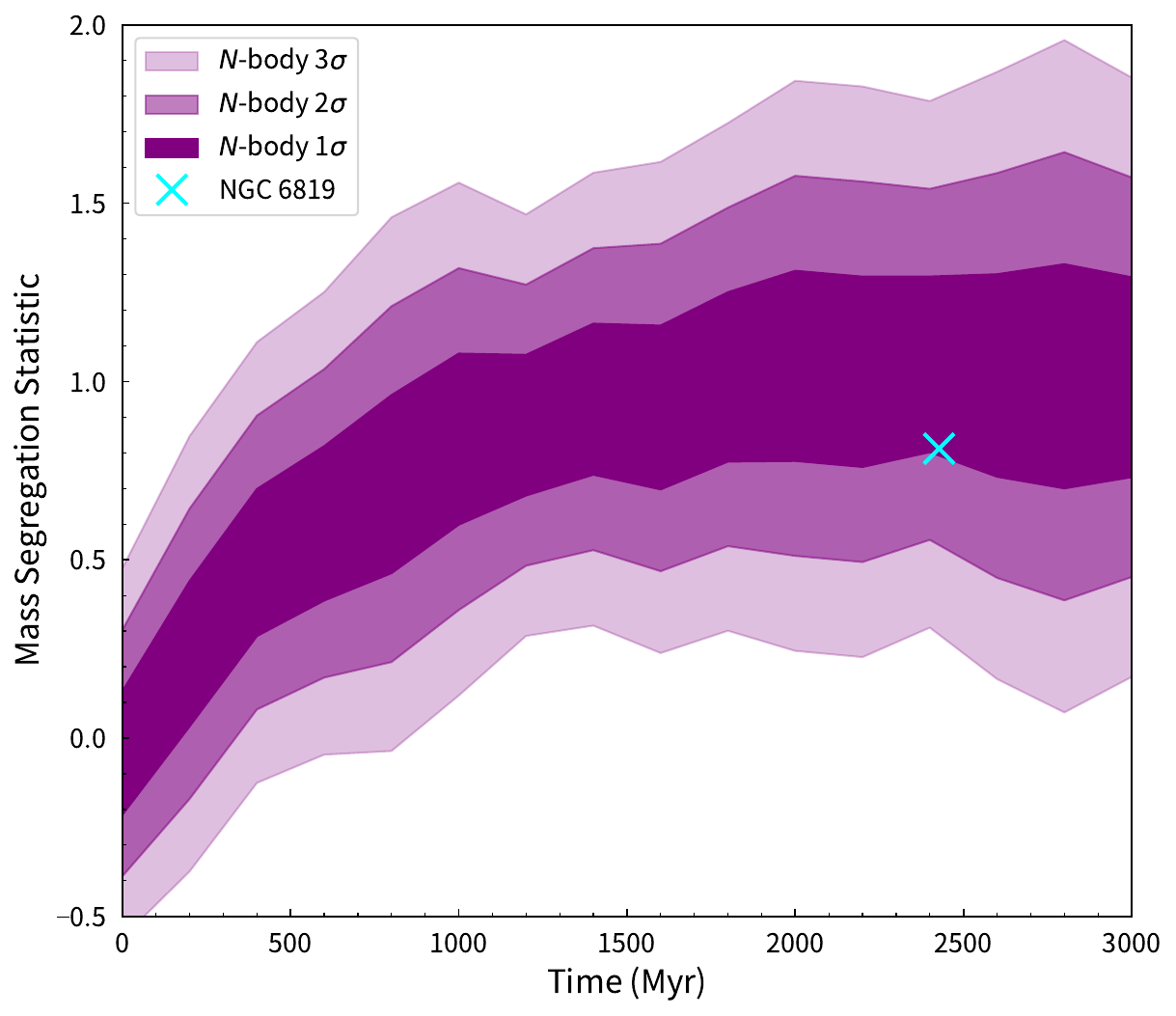}
\caption{Mass segregation statistic of the NGC 6819 model as a function of time for stars that would reside in our primary sample. The purple bands show the $1\sigma$, $2\sigma$, and $3\sigma$ bounds of the combined NGC 6819 model, in 200 Myr bins. The cyan ``X'' marks the calculated mass segregation statistic at the age of NGC 6819 found from our observational analysis. A positive mass segregation statistic indicates the binaries being shifted closer to the cluster center compared to the single stars.
\label{fig:mstat}}
\end{figure}

In Figure~\ref{fig:surfdensity}, we compare the (projected) surface density profile of the observations with that of the model at the age of NGC 6819.  For both the observations and the model, we include only stars within our primary sample (see Section~\ref{sec:primary_sample}).  For the model we select snapshots with ages within 3$\sigma$ of the median age of NGC 6819 found by \citet{2024ApJ...962...41C}; specifically we include snapshots with ages between 2414.5 Myr and 2440 Myr.  Then, for each snapshot in each of the individual simulations, we construct a surface density profile following the method of \citet{2013ApJ...779...30G, 2013AJ....145....8G, 2015ApJ...805...11G}, where we attempt to project each simulation on the plane of the sky (perpendicular to the line-of-sight) in order to compare the snapshot to the true cluster.  We use this projected view for all comparisons of the model with observations that depend on a radial component.  We then show the extent of all the surface density profiles for all realizations of the model cluster in the purple shaded region of Figure~\ref{fig:surfdensity}.   The region occupied by the NGC 6819 model comfortably includes the observed surface density profile of the true cluster.

Turning to the binaries at the age of NGC 6819, in Figure~\ref{fig:binaryfrac} we compare the observed (black) and simulated (cyan) binary fractions as a function of radius from the cluster center.  The model agrees well with the observations; a $\chi^2$ test shows that we cannot distinguish statistically between the model and observations (with a $p$-value of \chisquaredpval) in binary fraction vs. radius (for the primary sample and for $q > 0.5$).  

In Figures~\ref{fig:NbodyCDFbinaryvssingle} and \ref{fig:NbodymassCDFs} we perform a similar analysis for the NGC 6819 model as for the observations (which are shown in Figure~\ref{fig:CDFbinaryvssingle} and \ref{fig:massCDFs}).  For the simulations we combine all snapshots for all models that agree with the observed age of NGC 6819 into a single dataset from which to construct the CDFs; this provides a higher ``signal-to-noise'' analysis only possible with a set of $N$-body simulations.  The solar-type binaries in the model are mass segregated with respect to the single stars.  Furthermore, both the single stars and binaries, respectively, in the model show strong evidence for mass segregation (where the more massive samples have radial distributions that are shifted toward the cluster center with respect to lower mass samples).  This result mirrors those of the observations.

Finally, we investigate the time evolution of mass segregation of the binaries in Figure~\ref{fig:mstat}.  We follow a similar procedure to \citet{2016ApJ...833..252A} to generate a ``mass segregation statistic'' (similar to their ``$A^+$'' indicator) that is defined as the area between the two curves in a CDF plot of the radial distribution of the primary sample single stars compared to the primary sample binaries with $q > 0.5$ (e.g., the samples shown in Figure~\ref{fig:CDFbinaryvssingle} for the observations and Figure~\ref{fig:NbodyCDFbinaryvssingle} for the NGC 6819 model).  For both the observations and the simulations, we calculate the mass segregation statistic with respect to distance measured in parsecs.  A positive mass segregation statistic indicates that the radial distribution of the binaries is shifted closer to the cluster center than that of the single stars. 

In Figure~\ref{fig:mstat}, we show this mass segregation statistic from the start of the NGC 6819 model out to 3000 Myr in the purple region, and compare to the observed mass segregation statistic at the cluster age (shown in the cyan ``X'').  For the model, we measure the mass segregation statistic in each individual simulation at each snapshot, then take 200 Myr bins and show the $1\sigma$, $2\sigma$, and $3\sigma$ widths around the mean within each bin in the purple band.  In general, we see that the NGC 6819 model starts with essentially no measurable mass segregation (by design), and then as the cluster evolves dynamically, the mass segregation of the binaries increases.  At the age of NGC 6819, the model and observations agree within 1$\sigma$.  Interestingly, there is a large spread in the mass segregation statistic across the multiple $N$-body simulations that comprise our NGC 6819 model over all time.  The spread increases with time, due to the loss of stars (and therefore the degradation of the statistical power to distinguish the single and binary populations).  We return to this feature in Section~\ref{sec:discussion}.

\section{Discussion and Conclusions}\label{sec:discussion}

Through our analysis above, using a sample of only confident kinematic, spatial and photometric cluster members, we verify that NGC 6819 is mass segregated.  These results confirm findings in the literature \citep[e.g.,][]{Kalirai_et_al.,Kang_&_Ann,Yang_et_al.2013, Karatas_et_al.2023} using somewhat different samples and different techniques.  Importantly, here for the first time we also find that the binary stars show strong evidence of mass segregation, both with respect to the single stars and within the binary population itself.

We find that the radial distribution of the binaries is shifted significantly toward the cluster center with respect to the single stars, resulting in a binary fraction within NGC 6819 that decreases as a function of increasing distance from the cluster center (Figures \ref{fig:binaryfrac} and \ref{fig:CDFbinaryvssingle}).  Moreover, the binaries are more centrally concentrated than the single stars.  

Within our primary sample, the average mass of the binary stars is 1.75\ $M_\odot$, and the average mass of the single stars is 1.09\ $M_\odot$.   The timescale for objects of these masses to segregate (Equation~\ref{eq2}) is roughly $\sim$5-8 times less than the cluster age.  Thus, we interpret this difference in radial distributions to be a result of mass segregation processes.  

Additionally, we find strong evidence for mass segregation within the binary population itself, as shown in the left panel of Figure \ref{fig:massCDFs}.  Increasingly massive binaries show increasingly centrally concentrated radial distributions, a confirmation that the binaries are mass segregated in NGC 6819.

These observational results are also supported by our NGC 6819 $N$-body model (Section~\ref{sec:Nbodyresults}).  We find that a model which reproduces the observed surface density profile and solar-type binary fraction of the true cluster at the age of NGC 6819 also displays strong evidence for mass segregation both within the single and binary populations themselves and when comparing the binary and single populations to each other.  

Our NGC 6819 $N$-body model also suggests that the true cluster was likely born with a lower binary fraction than is observed (today) in the Galactic field (by $\sim$30\%).  Today, NGC 6819 has a binary fraction for solar-type stars that is roughly consistent with the field \citep{Hole_et_al._2009, 2024ApJ...962...41C}.  The binary fraction increases over time in these models due to the preferential evaporation of single stars (as has also been noted previously in the literature for other star cluster $N$-body models, e.g., \citealt{2007ApJ...665..707H}). 

Interestingly, when investigating the development of mass segregation over time in the NGC 6819 model (Figure~\ref{fig:mstat}) we see a rather large scatter in the mass segregation statistic.  The amount of scatter increases with time, due to the loss of both binary and single stars.  Importantly, this may provide some insight to explain the observational results discussed in Section~\ref{sec:intro}, where some clusters of similar age to NGC 6819 are observed to be mass segregated and others are not.  At an age of $\sim$2.5 Gyr, one would need a cluster with a mass of order $\sim2\times10^{5}M_\odot$ (with a half-mass radius of $\sim$5pc, and a mean mass of 0.5M$_\odot$) to achieve a relaxation time of order the cluster age.  This is much larger than typical open clusters in our Galaxy, and therefore one would naively assume that the Galactic open clusters that have survived to the age of NGC 6819 would be mass segregated.  The scatter in the $N$-body model shows that stochasticity and small sample sizes may be at least part of the explanation between these differing observations.     

Mass segregation of binaries as compared to single stars has also been reported in a number of other star clusters over a wide range in age, from globular clusters \citep{2012A&A...540A..16M}, to old open clusters like M67 and NGC 188 \citep{2015AJ....150...97G, Geller_et_al._2008} to younger open clusters like Pleiades, Praesepe, and M35 \citep{1998A&A...333..897R, 2024ApJ...962L...9M}.   With this work, NGC 6819 now provides a snapshot at an intermediate age to further study the effects of mass segregation of binary stars.

\begin{acknowledgments}
This material is based upon work supported by the National Science Foundation (NSF) under Grant No. AST-2149425, a
Research Experiences for Undergraduates (REU) grant, and under the NSF AAG Grant No.\ AST-2107738. The material contained in this document is also based upon work supported by a National Aeronautics and Space Administration (NASA) grant awarded to the Illinois/NASA Space Grant Consortium. Any opinions,
findings, and conclusions or recommendations expressed in this material are those of the author(s) and do not necessarily
reflect the views of the NSF or NASA. This research was supported in part through the computational resources and staff contributions provided for the Quest high performance computing facility at Northwestern University
which is jointly supported by the Office of the Provost, the Office for Research, and Northwestern University
Information Technology. We would also like to thank Elizabeth Jefferey, Roger Cohen,
Elliot Robinson, and the rest of the BASE-9 group for their insights and support.
\end{acknowledgments}

\bibliography{main}{}
\bibliographystyle{aasjournal}

\end{document}